# Staircase-like transfer characteristics in multilayer MoS$_2$ field-effect transistors

Takuya Ohoka[1] and Ryo Nouchi[1,2]

[1] Department of Physics and Electronics, Osaka Prefecture University, Sakai 599-8570, Japan
[2] PRESTO, Japan Science and Technology Agency, Kawaguchi 332-0012, Japan

E-mail: r-nouchi@pe.osakafu-u.ac.jp

**Abstract**

Layered semiconductors, such as MoS$_2$, have attracted interest as channel materials for post-silicon and beyond-CMOS electronics. Much attention has been devoted to the monolayer limit, but the monolayer channel is not necessarily advantageous in terms of the performance of field-effect transistors (FETs). Therefore, it is important to investigate the characteristics of FETs that have multilayer channels. Here, we report the staircase-like transfer characteristics of FETs with exfoliated multilayer MoS$_2$ flakes. Atomic force microscope characterizations reveal that the presence of thinner terraces at the edges of the flakes accompanies the staircase-like characteristics. The anomalous staircase-like characteristics are ascribable to a difference in threshold-voltage shift by charge transfer from surface adsorbates between the channel center and the thinner terrace at the edge. This study reveals the importance of the uniformity of channel thickness.

## 1. Introduction

With regard to the continued efforts to realize high-performance electronic devices, miniaturization of field-effect transistors (FETs) incorporated in integrated circuits has been pursued. However, a reduction in FET dimension deteriorates its performance due to the so-called short-channel effect [1-3], preventing further miniaturization. The short-channel effect makes it difficult for the FET channel to be depleted by the gate voltage. As a result, deteriorative effects occur upon miniaturization, such as an increase in power consumption due to an increase in off current. To suppress the short-channel effect, the FET channels should be sufficiently thin to be fully depleted by the gate electric field. Therefore, layered semiconductors typified by transition metal dichalcogenides (TMDCs), such as MoS$_2$, have attracted interest because ultrathin channels can be obtained by exfoliation of such layered semiconductors [4-9].

Considerable research on layered semiconductors has been devoted to the monolayer limit. However, a dependence of FET performance on the number of layers has shown that the monolayer channel is not advantageous in terms of field-effect mobilities [10,11]. A band gap of semiconducting TMDCs, such as MoS$_2$, increases as the number of layers is reduced [12], leading to an increase in the Schottky barrier height at metal contacts to a few-layer TMDCs [10]. Furthermore, additional layers screen the Coulomb potential of charged impurities to reduce the scattering probability [11]. The possibility of the weakening of the interaction between electrons and phonons is also discussed [11]. All of these features indicate that the performance of FETs is higher with multilayered channels. Therefore, an investigation on FET characteristics based on multilayer semiconducting TMDCs is also of critical importance.

In this article, we report the anomalous characteristics frequently observed in FETs with exfoliated multilayer MoS$_2$

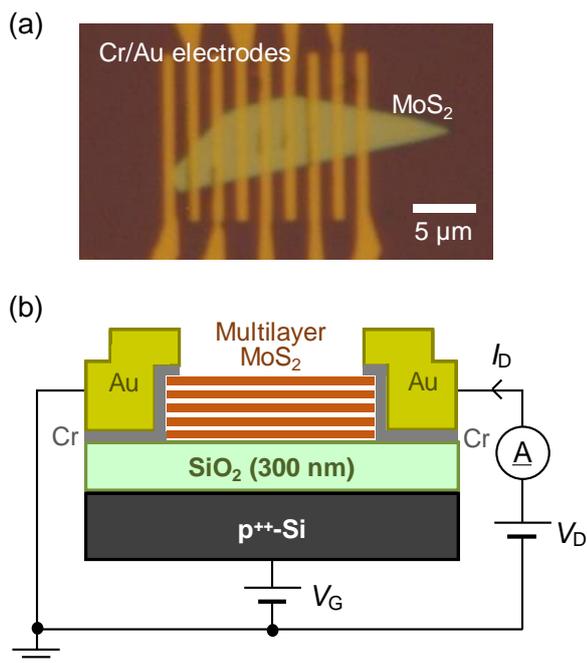

**Figure 1**. Structure of fabricated FETs based on exfoliated multilayer MoS$_2$. (a) Optical micrograph. (b) Schematic diagram.

flakes. Specifically, they display transfer characteristics (dependence of the drain current, $I_D$, on the gate voltage, $V_G$) with a two-tiered staircase-like structure, which is completely different from commonly observed structures. A careful inspection of the flake structure by atomic force microscopy (AFM) shows that the presence of a thinner terrace at the edges of the MoS$_2$ flakes accompanies the staircase-like transfer characteristics. The anomalous characteristics explain that a smaller $V_G$ is required to build up a channel in the thinner terraces near the edges than that in the thicker part around the center of the flake. The difference in the threshold voltage, $V_{th}$, between the two distinct parts within the flake is ascribable to a difference in $V_{th}$ shift by charge transfer from surface adsorbates between the two parts. These considerations are further supported by the investigation of the effect of thermal annealing treatment. Our results are based on thick MoS$_2$ flakes, but the conclusions also hold for ultrathin flakes. This study clearly shows the importance of the uniformity in channel thickness.

## 2. Experimental

MoS$_2$ flakes obtained by mechanical exfoliation from a natural crystal (SPI Supplies) were transferred to a highly doped silicon substrate (p$^+$-Si) that has a 300-nm-thick thermal oxide layer. Source/drain electrodes of a 50-nm-thick Au film with a 1-nm-thick Cr adhesion layer were fabricated by standard electron-beam lithography processes. Figures 1(a) and 1(b) show an optical micrograph and a schematic diagram of fabricated devices, respectively. All the channel lengths were designed to be 1 µm. Electrical characterizations were conducted with a semiconductor device analyzer (Keysight, B1500A) at room temperature. Transfer characteristics were measured with the drain voltage, $V_D$, of 0.1 V and $V_G$ swept from −80 V to +80 V in 0.5 V steps. First, the FETs were measured under ambient conditions, and then, in order to remove surface adsorbates, they were heated at 200 °C for 2 h after thoroughly purging the ambient air in the measurement chamber with N$_2$ gas; the FETs were measured again without exposure to the ambience after cooling down to room temperature. Finally, we acquired surface topographic images of the measured FETs through AFM by the dynamic force mode (Hitachi High-Technologies, AFM5200S).

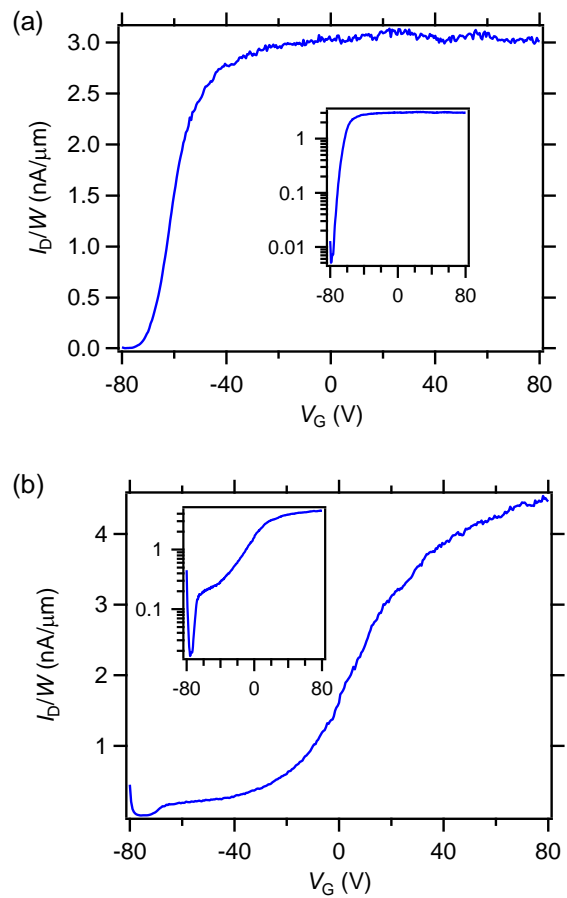

**Figure 2**. Transfer characteristics of as-fabricated FETs based on exfoliated multilayer MoS$_2$ measured in ambient air. (a) Commonly observed one-tiered characteristics. (b) Anomalous two-tiered staircase-like characteristics.

## 3. Results and discussion

Figure 2 shows transfer characteristics normalized by the channel width, $W$, of the fabricated multilayer MoS$_2$ FETs.

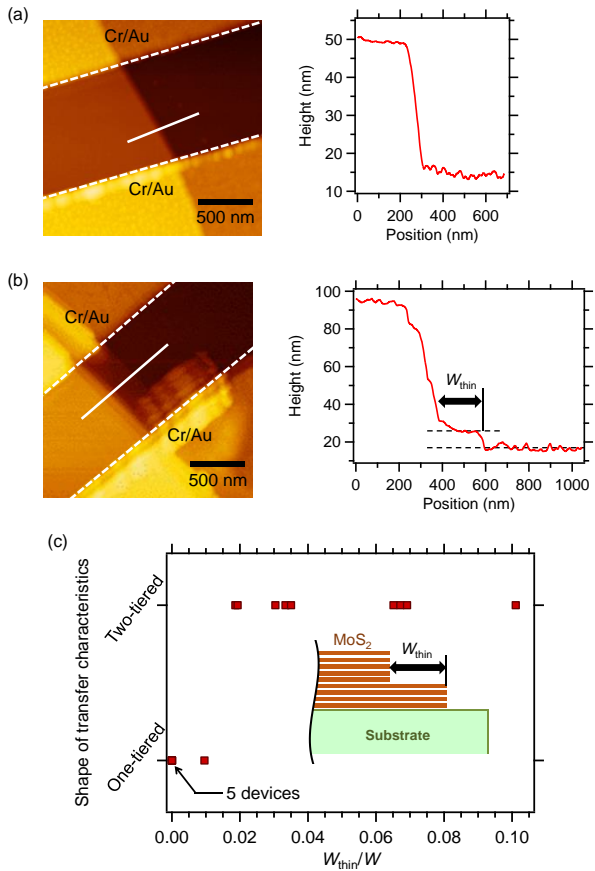

**Figure 3.** Surface topographic characterizations of the channels of exfoliated multilayer MoS$_2$. (a, b) AFM images (left) and height profiles along solid white lines indicated in the AFM images (right) of the FETs identical to (a) figure 2(a) and (b) figure 2(b). (c) The correlation with the shape of the transfer characteristics and the width of the thinner part within the channel. The inset is a schematic representation of the width of the thinner part, $W_{thin}$.

The $I_D$ shows a quasi-saturation behavior, which has been understood as a characteristic of a Schottky FET where modulation of the Schottky barrier width dominates the change in $I_D$ [13]. We confirmed that our devices showed a similar behavior as shown in figure 2(a). In addition to the commonly observed characteristics, we also frequently observed the characteristics as shown in figure 2(b). The $I_D$ increased again when $V_G$ reached ca. $V_{th} + 50$ (V), resulting in transfer characteristic with a two-tiered staircase-like structure. The staircase-like characteristics are clearly discernible in the semi-log plot [the inset of figure 2(b)]. To the contrary, the characteristics shown in figure 2(a) does not display a staircase-like curve, even in the semi-log plot shown in the inset.

The two-tiered staircase-like transfer characteristics shown in figure 2(b) are determined by considering a combination of two distinct channels with a different $V_{th}$ value. It is known that the conducting channel of MoS$_2$ FETs is formed initially at the edges and then expands to the entire flake [14]. This has been understood by the presence of topologically trivial electronic states localized at the edges [15,16], which are first populated by the gating [14]. Besides, the finite channel width of actual FETs also leads to the channel formation initially at the edges. An ideal FET is generally characterized as a parallel-plate capacitor with infinite plates, where the gate electric field is perpendicular to the plates. However, the gate field becomes non-perpendicular near the edges of finite plates, which is called the fringing field [17,18]. Thus, an effective magnitude of $V_G$ is larger at the channel edges, and the channel should be formed initially at the edges. Therefore, the $V_{th}$ value should be lower at the edges than the center of the channel.

The considerations above suggest that the first tier in the anomalous transfer characteristics in figure 2(b) originates from the edge-related channel possessing a lower $V_{th}$. If the staircase-like characteristics appear as a consequence of the edge states and the fringing field, the characteristics should also be observed with a monolayer channel. However, this has not been the case so far [6,13,14,19-23]. Therefore, the anomalous characteristics would arise from a difference possibly in edge structure between the multilayer channel and the monolayer channel.

Figure 3 shows AFM topographic images of the devices identical to the ones measured in figure 2, where figures 3(a) and 3(b) correspond to the device in figures 2(a) and 2(b), respectively. As clearly discernible in figure 3(b), the flake thickness near the edge was found to be thinner than that at the center of the channel. However, such a step-terrace structure was not observed in figure 3(a), which is naturally the case with monolayer channels. We characterized 15 FETs in total, and found that the presence of the thinner part within the channel always accompanied the staircase-like transfer characteristics. Figure 3(c) shows the correlation between whether a staircase-like structure is seen in the transfer characteristics and the width of the thinner part around the edge. We judged that the transfer characteristics possess a staircase-like structure if their transconductance curves contain a local minimum in addition to local maxima. The staircase-like characteristics are less likely to be seen when the thinner part is narrow or indiscernible through AFM observation. These findings indicate that the presence of a thinner part at the flake edges is necessary to observe such staircase-like characteristics.

One possible mechanism that relates the thickness variation to the difference in $V_{th}$ is the charge transfer from surface adsorbates. In the case of back-gated FETs, such as those investigated in this study, the channel is located away from the surface with the adsorbates by the flake thickness. Charge carriers doped from the adsorbates affect the charge-carrier concentration within the channel, though the change in carrier concentration by surface doping is reduced by a screening effect as the distance from the surface adsorbates becomes longer. The change that remains at the channel should be higher with a smaller flake thickness, as schematically shown in figure 4(a). Therefore, $V_{th}$ is expected to shift with the film thickness in the presence of charge transfer from surface adsorbates.

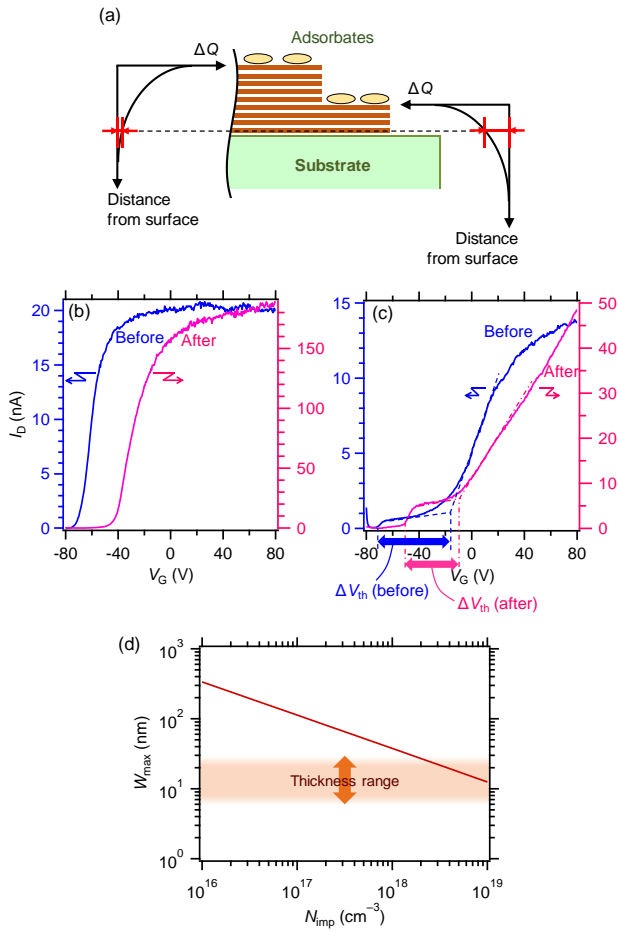

**Figure 4.** Effect of charge transfer from surface adsorbates. (a) A schematic diagram of the effect of surface charge transfer on carrier concentrations in the channel. The $V_{th}$ shift caused by the charge transfer is expressed as $\Delta Q/C_0$, where $\Delta Q$ is the change in charge concentration and $C_0$ is the gate dielectric capacitance per unit area. (b, c) Transfer characteristics of FETs based on exfoliated multilayer $MoS_2$. The pink and blue lines indicate the characteristics measured before and after the annealing, respectively. The FETs of (b) and (c) are identical to those in figures 2(a) and 2(b), respectively. (d) Calculated relationship between $W_{max}$ and $N_{imp}$, together with the thickness range of the thinner part.

If the difference in $V_{th}$ between the first and second tiers in the staircase-like transfer characteristics, $\Delta V_{th}$, is caused by the influence of the surface adsorbates, the $\Delta V_{th}$ value should be reduced by removing the adsorbates. We performed thermal annealing of the devices at 200 °C for 2 h in a $N_2$ environment and measured them again in the same $N_2$ environment after cooling down to room temperature. Figures 4(b) and 4(c) compare the transfer characteristics before and after the thermal treatment, where the characteristics measured before the treatment are identical to those shown in figures 2(a) and 2(b), respectively. In both devices, the transfer characteristics shifted towards the positive-$V_G$ direction by the thermal treatment. This shift suggests that electrons are transferred from the surface adsorbates. While air molecules, such as oxygen and water, are known to induce hole doping [24,25], substances that induce electron doping are unknown at present. We estimate that the electron doping was induced by resist residues deposited during the lithographic processes, which has been suggested to induce electron doping [26]. In the case of the two-tiered staircase-like transfer characteristics shown in figure 4(c), the amount of the $V_{th}$ shift by the thermal treatment differs between the first tier (19.7 V) and the second tier (8.7 V). The $\Delta V_{th}$ value decreased from 55.4 V to 44.4 V by the thermal treatment as expected from the mechanism based on charge transfer from surface adsorbates.

However, when the film thickness is too large, the carrier concentration in the channel cannot be modulated by the surface charge transfer. Here, the maximum depletion width, $W_{max}$, is considered as an index of the thickness range affected by the charge transfer. $W_{max}$ can be expressed as [27]

$$W_{max} = \sqrt{\frac{4k_B T \varepsilon_{MoS_2} \varepsilon_0}{q^2 N_{imp}}} ln \frac{N_{imp}}{n_i},$$

where $k_B$ is the Boltzmann constant, $T$ is the absolute temperature, $\varepsilon_{MoS_2}$ is the dielectric constant of $MoS_2$, $\varepsilon_0$ is the permittivity of vacuum, $q$ is the elementary charge, $N_{imp}$ is the density of ionized impurities, and $n_i$ is the intrinsic carrier density. Among the material parameters, $\varepsilon_{MoS_2} \approx 11$ [28] and $n_i = 1.6 \times 10^8$ cm$^{-3}$ at room temperature [27,29]. $N_{imp}$ has been reported to vary from $10^{16}$ to $10^{19}$ cm$^{-3}$ in $MoS_2$ [27,30]. $W_{max}$ corresponding to the $N_{imp}$ range was calculated to be from 336 to 13 nm. The thicknesses of the thinner part ranges from 6 to 30 nm in the $MoS_2$ channels that showed two-tiered transfer characteristics. A part of the thickness range is within the $W_{max}$ range calculated above (figure 4(d)). However, the centroid of electrical current distribution in top-contact bottom-gate FETs with a layered semiconductor channel has been shown to migrate from near the bottom surface to the top surface with increasing $V_G$ [31]. This consideration, based on a resistor-network model, indicates that the $I_D$ level can be affected even when the thinner part is thicker than $W_{max}$. Therefore, the charge transfer from surface adsorbates should exert a sufficiently large effect on $I_D$ even with a relatively thick terrace at the edges.

It should be noted that the thickness range of flakes used in this study is not sufficiently thin for suppression of short-channel effects. However, as schematically indicated in figure 4(a), the effect of the surface charge transfer decays exponentially (or in a similar manner) with increasing thickness. This indicates that the difference in $\Delta Q$ between a $n$-layer flake and a $(n + 1)$-layer flake becomes larger for smaller $n$, where $n$ is a natural number. Therefore, $\Delta V_{th}$ induced by the thickness inhomogeneity is expected to be significant also for the ultrathin regime.

## 4. Conclusions

In conclusion, we report a two-tiered staircase-like shape in transfer characteristics of FETs based on an exfoliated multilayer MoS$_2$ flake, which is completely different from the commonly observed one-tiered characteristics. AFM inspections on the flake topography revealed that the presence of a thinner part at the edges of the flake accompanies the staircase-like characteristics. The two-tiered shape of the characteristics, which indicates the presence of two distinct regions with different $V_{th}$ values, can be explained by the channel formation initially at the thinner part at the edges because a shift in $V_{th}$ caused by surface charge transfer is larger in a thinner flake. The staircase-like transfer characteristics deteriorate switching performance of FETs, which should be avoided. Thus, the present finding clearly shows the importance of the uniformity in thickness of multilayered channels, which is significant not only for the thick flakes studied here but also for ultrathin flakes. Although research on layered semiconductors has mostly focused on the monolayer limit, the performance of FETs based on layered semiconductors has been reported to be higher in multilayered channels than monolayered channels [10,11]. Thus, the anomalous characteristics unveiled in this study, which is observable only with multilayered channels, are of critical importance for the analyses of high-performance FETs based on layered semiconductors.

## Acknowledgements

This work was supported by JSPS KAKENHI Grant Numbers JP17H01040 and JP19H02561; and JST, PRESTO Grant Number JPMJPR17S6.

## Data availability

The data that support the findings of this study are available from the corresponding author upon reasonable request.